\documentclass[10pt,aps,prb,preprint]{revtex4}   

\usepackage{amsmath}    
\usepackage{amsfonts}  
\usepackage{amssymb}
\usepackage{graphicx}   

\usepackage[latin1]{inputenc}     
\usepackage[T1]{fontenc}
\graphicspath{{plots/}{../images/}}

\usepackage{float}   

\newtheorem{Theoreme}{THEOREM}

\newtheorem{Proposition}{Proposition}
\newtheorem{Remarque}{Remark}
\newtheorem{Conjecture}{Conjecture}

\begin{document}

\title{External and internal wave functions:\\de Broglie's double-solution theory~?}
\author{Michel Gondran}
 \affiliation{Académie Européenne Interdisciplinaire des Sciences, Paris, France}
 \email{michel.gondran@polytechnique.org}   
 \author{Alexandre Gondran}
\affiliation{\'Ecole Nationale de l'Aviation Civile, 31000 Toulouse,
France}
 \email{alexandre.gondran@enac.fr}   

\begin{abstract}

We propose an interpretative framework for quantum mechanics corresponding to the specifications of Louis de Broglie's double-solution theory. The principle is to decompose the evolution of a quantum system into two wave functions: an external wave function corresponding to the evolution of its center of mass and an internal wave function corresponding to the evolution of its internal variables in the center-of-mass system. Mathematical decomposition is only possible in certain cases because there are many interactions linking these two parts. In addition, these two wave functions will have different meanings and interpretations.

The external wave function "pilots" the center of mass of the quantum system: it corresponds to the Broglie pilot wave. When the Planck constant tends to zero, it results mathematically from the convergence of the square of the module and the phase of the external wave function to a density and a classical action verifying the Hamilton-Jacobi statistical equations. This interpretation explains all the measurement results, namely those yielded by interference, spin measurement (Stern and Gerlach) and non-locality (EPR-B) experiments. 

For the internal wave function, several interpretations are possible~: the one of the pilot wave can be applied in cascade to the internal wave function. 
However, the interpretation proposed by Erwin Schrödinger at the Solvay Congress in 1927 and restricted to the internal wave function is also possible.
For Schrödinger, the particles are extended and the square of the module of the (internal) wave function of an electron corresponds to the density of its charge in space. 
We present  many arguments in favour of this interpretation, which like the pilot wave interpretation is realistic and deterministic. 

Finally, we will see that this double interpretation serves as a frame of reference by which to better understand the debates on the interpretation of quantum mechanics and to review the relationships between gravity and quantum mechanics.
\end{abstract}

\maketitle
\section{Introduction}

The fathers of quantum mechanics strongly disagreed at the 1927 Solvay Congress regarding the interpretation of the wave function of a quantum system. And the debate continues today.

As soon as his equation was defined, Schrödinger dreamed of the possibility of building a non-dispersive wave packet completely representing the
particle. This was how he
introduced the coherent states of the harmonic oscillator in 1926~\cite{Schrodinger1926}:
\textit{``Our wave packet always remains grouped, and does not spread over an increasingly large space over time, as do, for example, wave packets that we are used to considering in optics.''}
But he fails to address the problem of the hydrogen atom by finishing his article in this way:
\textit{``It is certain that it is possible to construct by a process quite similar to the previous one, wave packets gravitating on Kepler ellipses at a large number of quanta and forming the wave image of the electron of a hydrogen atom; but in this case the difficulties of calculation will be much greater than in the particularly simple example that we have treated here and which from this point of view is almost a course exercise''}. He took up this interpretation~\cite{Schrodinger1952} again in 1952 in a debate with the Copenhagen school.
 
\bigskip

For de Broglie, the double-solution theory is the true interpretation, which he outlined~\cite{deBroglie1927} in 1926 and sought to demonstrate his whole life~\cite{deBroglie1956,deBroglie1971,deBroglie1977},
the pilot wave he presented at the Solvay Congress in 1927 being only a by-product: 
\textit{``I was introducing,
under the name of "double-solution theory" the idea
that \textbf{we had to distinguish between two distinct solutions, but
intimately related to the wave equation}, one that I called
the $u$ wave being a real and non-standard physical wave
with a local accident defining the particle and
represented by one singularity, the other, Schrödinger's $\Psi$ wave , normalizable and devoid of singularity, which would not be
that a representation of probabilities~\cite{deBroglie1971}}''.

\bigskip

In the 2017 special issue of the \textit{Annales de la Fondation de Broglie}, we find a synthesis of recent work on Broglie's double-solution and its history over the past 90 years~\cite{Fargues2017,Colin2017,Robert2017} since de Broglie presented the first ideas in 1927. Non-linear physics occupies an important part of this work as well as the concept of soliton, described as a  \emph{``singular and persistent object, [a] materialization in wave form of the corpuscle concept''} which allows to reconsider the program of the double solution and offers promising prospects for the reconciliation of quantum theory with realism~\cite{Colin2017,Matzkin2017}. The addition of non-linear terms to the Schrödinger equation is widely studied.
Of particular note is Thomas Durt's article~\cite{Durt2017b} which defines a very interesting double solution \textit{à la de Broglie} for Schrödinger-Newton's non-linear equation.

The approach we adopt in our article differs from that of the special issue in that we do not introduce any non-linearity,
coherent states showing that the concept of soliton also appears in linear physics. 
The double solution we present is more a distinction of scale (external/internal) and interpretation than a search for an underlying non-linear wave equation~\cite{Robert2017}.

\bigskip

In his 1954 Nobel speech, Born~\cite{Born1955} recalls his approach to defining the statistical interpretation of quantum mechanics~: \textit{``It was again Einstein's idea that guided me.
He had tried to make the duality of the
waves and particles - light quanta or
photons - comprehensible by considering the square of the light wave amplitudes as the probability density for the presence of photons. This idea could immediately extend to the function $\psi $ : $\mid\psi\mid^2 $
should be the probability density for the presence of electrons (or other particles). It was easy to
assert this. But how can it be demonstrated~?
The atomic collision processes  made it possible.''}

\bigskip

In contrast to de Broglie
and Schrödinger's search for physical images, Heisenberg presents a formal framework for
quantum theory, theorized as a system of concepts~\cite{Heisenberg1927}
\textit{``The new concept system at the same time yield the
intuitive content of the new theory. From an intuitive theory
in this sense we must therefore only ask that it be without
contradiction and that it be able to predict unambiguously the
results of every conceivable experiment in his field.
Quantum mechanics will be in this sense an intuitive and complete theory 
of mechanical processes.''} Thus, Heisenberg only keeps the minimal, non-contradictory condition.

For Niels Bohr, the theory
is based on the principle of complementarity~\cite{Bohr1927} with which he
hopes to\textit{``contribute to by reconciling the apparently
contradictory conceptions defended by different
physicists''}. He argues that it is
possible to express the essence of the theory using the
\textbf{"quantum postulate"}. [...] \textit{``This quantum postulate implies that any
observation of atomic phenomena will involve an interaction
 with the agencies of observation; therefore, an independent reality in the ordinary physical sense can neither be ascribed to the phenomena nor to the agencies of observation. [...] we must regard them as complementary, but
mutually exclusive features of our representation of the
experimental findings.''}

\bigskip

Einstein summed up this debate well in one of his final texts (1953), \textit{Elementary Considerations on the Interpretation of the Foundations of Quantum Mechanics}
in homage to Max Born: \textit{``The fact that the Schrödinger equation combined with the
Born interpretation does not lead to a description of the
"real state" of a single system, naturally gives rise to
a search for a theory which is free of this
limitation. So far there have been two attempts in this direction, which share the features that they maintain the Schrödinger equation, and give up the Born interpretation. \textbf{The first effort} goes back to de Broglie and has been pursued further by Bohm with great perspicacity [...] \textbf{The second attempt}, which aims at achieving a "real description" of an individual system, based on the Schrödinger equation, has been made by Schrödinger
himself. Briefly, his ideas are as follows. \textbf{The $\psi$-function itself represents reality,}
and does not stand in need of the Born interpretation....[...] From the
previous considerations, it follows that the only acceptable interpretation
of Schrödinger's equation up to now is 
the statistical interpretation given by Born. However, it
does not give the "real description" of the individual system, but
only statistical statements related to sets
of systems.''}~\cite{Einstein1953}

\bigskip

We propose here an interpretation of quantum mechanics that corresponds to the double-solution theory sought by Louis de Broglie. This interpretative framework is here limited to mass and non-relativistic particles.

This interpretation follows our work on the theory of double preparation~\cite{Gondran2014c} where we show that there are two types of interpretation following the preparation of the quantum system. The double solution provides an explanation for this double preparation.
The basic idea is to study the evolution of a quantum system from the evolution of its center of mass (external evolution) as well as its internal evolution. The use of the center-of-mass wave is not new,
and has been mentioned very often by many authors. For example, it is very well explained in the book \emph{Atomic Interferometry} by Baudon and Robert~\cite{Baudon2004}: \textit{``In the free evolution of an atom or molecule, the external motion (corresponding to the motion of the mass center \textbf{R}) and the wave associated with it play a separate role. Indeed, because of the separation of the Hamiltonian from the system (in the absence of any external interaction) in the form: $H=T + H_{int}$ where $T=-\frac{\hbar^2}{2m} \vartriangle_R $
is the kinetic energy operator and $ H_{int}$ the part of the Hamiltonian that involves only variables other than \textbf{R}, there are system states whose wave function has the form $\Psi(R,t) \Phi_{int}$, where $\Psi(R,t)$ and $\Phi_{int}$
are proper states of T and $ H_{int}$ respectively.  It is on the wave function $\Psi $ of the external motion, which corresponds to a state of the continuum, that atomic interferometry will be carried out.''} A similar point of view is adopted by Claude Cohen-Tannoudji in the preface to this book~\cite{Baudon2004}: \textit{``A de Broglie wave is also associated with the movement of the center of mass of a more complex quantum system, such as an atom or molecule, composed of several protons, neutrons and electrons. The wavelength of the de Broglie wave associated with an object of mass $M$ and velocity $v$ is inversely proportional to the product $Mv$.''}

\bigskip

The approach is the same as in classical mechanics: it consists in studying the evolution of a system from its external variables such as the center of mass and its velocity, which correspond to the global motion of the quantum system, and the internal variables, which correspond to its motion in the reference frame of the center of mass. We study how the wave function of the system can be decomposed into two wave functions: the wave function of its center of mass (external evolution) and the internal wave function. These two wave functions, which correspond to a separation of the energy spectrum into a continuous spectrum and a discrete spectrum, are of a different nature and will have different interpretations, which logically leads to the double-solution theory sought by Louis de Broglie.

First, we show that the external wave function corresponds to the interpretation of the de Broglie-Bohm "pilot wave" (dBB). Indeed, we demonstrate mathematically that, when we make the Planck constant h tend towards zero, the square of the module and the phase of the external wave function converge towards a density and a classical action verifying the Hamilton-Jacobi statistical equations. This interpretation of the external wave function by the de Broglie-Bohm pilot wave gives a physical explanation for the measurement results, both for diffraction and interference experiments as well as for spin measurements as in the Stern-Gerlach and EPR-B experiments. The reduction of the wave function of the quantum system is then controlled by the position of the center of mass at the time and position where the quantum system is captured.

For the internal wave function, several interpretations are possible. We will study three possibilities of interpretation~: that of dBB, that of Copenhagen and the interpretation that corresponds to that proposed by Erwin Schrödinger in 1926 and then at the Solvay Congress in 1927.
He argues that particles are extended and the square of the module of the (internal) wave function of an electron corresponds to the value of the distribution of its charge in space.
The interpretation of the internal wave function is therefore deterministic, and the double-solution theory is also a deterministic theory.

The plan of the article is as follows. 
In section~\ref{sect:decomposition}, we study how the wave function of a N-body quantum system is decomposed between its external and internal wave functions. We also present two non-quantum analogies to better understand the interaction between external and internal wave functions. 
In section~\ref{sect:interpretationExterne}, we demonstrate mathematically, by studying convergence towards classical mechanics, and experimentally, by simply explaining measurement in quantum mechanics, that the most plausible interpretation of the external wave function is that of de Broglie-Bohm "pilot wave". 
In section~\ref{sect:interpretationInterne}, we propose an interpretation of the internal function corresponding to the Schrödinger interpretation. 
Then in section~\ref{sect:experienceCruciale}, we propose a crucial experiment to validate the existence of external and internal wave functions.
Finally, in conclusion, we show that this double interpretation clarifies debates on the interpretation of quantum mechanics and to review the relationships between gravity and quantum mechanics.

\section{External and internal wave functions}
\label{sect:decomposition}

From the beginning of quantum mechanics, two types of variables have been distinguished for studying atomic or molecular dynamics: internal and external variables. The external variables concern the external dynamics of the atom, i.e. the movement of its center of mass and the orientation of the frame of reference linked to it. Internal variables describe, for example, the evolution of the structure of the atom or the molecule.

These internal and external degrees of freedom are not generally independent. The interactions between internal and external variables are indeed at the basis of the manipulation of atoms, in particular their cooling~\cite{Dalibard2006}. 

Depending on the experimental conditions, these interactions vary in size, making the decomposition of the total wave function into an external wave function and an internal wave function more or less approximated. Let's start by studying the case of a N-body system where this decomposition is accurate.

\subsection{Decomposition of a N-body system such as an atom or molecule}
Let us consider a system of N particles without spin, with masses $m_i$ and coordinates $\textbf{x}_i$, subjected to a linear potential field $V_i(\textbf{x}_i)=m_i \textbf{g} .\textbf{x}_i$, and to mutual interactions described by the potentials $U_{ij}(\textbf{x}_i -\textbf{x}_j)$. This quantum system is therefore described by the wave function $ \Psi^h(\textbf{x}_1, \textbf{x}_2,...,\textbf{x}_N, t) $, which satisfies the Schrödinger equation:
\begin{equation}\label{eq:schrodinger1d}
i\hslash \frac{\partial \Psi^h(\textbf{x}_1, \textbf{x}_2,.., \textbf{x}_N, t) }{\partial t}= H \Psi^h(\textbf{x}_1, \textbf{x}_2,..,\textbf{x}_N, t) 
\end{equation}
with the Hamiltonian
\begin{equation}\label{eq:hamiltonschrodinger1}
H= \Sigma_{i} (\frac{\textbf{p}_i ^{2}}{2m_i} +V_i(\textbf{x}_i))+ \Sigma_{ij} U_{ij}(\mathbf{x}_i -\textbf{x}_j)). 
\end{equation}
and the initial condition 
\begin{equation}\label{eq:schrodinger2}
\Psi^h (\mathbf{x}_1, \textbf{x}_2,.., \textbf{x}_N, 0)=\Psi^h_{0}(\mathbf{x}_1,\textbf{x}_2 , ..,\textbf{x}_N).
\end{equation}

The movement of these N particles is separated from the movement of their center of mass as in classical mechanics: Let $\textbf{x}_G=(\Sigma_{i} m_i \textbf{x}_i)/(\Sigma_i m_i)$ be the position of the center of mass, $\textbf{x}_i'= \textbf{x}_i -\textbf{x}_G$ the position of the particle i relative to the barycenter, $M= \Sigma_i m_i $ the total mass, $\textbf{x}'_G=(\Sigma_{i} m_i \textbf{x}'_i)/(\Sigma_i m_i)= \textbf{0}$ the internal coordinates of the center of mass. Then the Hamiltonian H is written according to the total impulses ($\textbf{p}_G= \Sigma_i \textbf{p}_i $) and relative impulses ($\textbf{p}'_i= \textbf{p}_i -m_i/M \textbf{p}_G $):
\begin{equation}\label{eq:hamiltonschrodinger2}
H=  \frac{\textbf{p}_G^{2}}{2M} + M \textbf{g} . \textbf{x}_G+  \Sigma_i \frac{\textbf{p}'^{2}}{2 m_i} + \Sigma_{ij}U_{ij}(\mathbf{x}'_i -\textbf{x}'_j) = H_{ext} + H_{int}. 
\end{equation}
and there is no interaction between internal and external variables.
 
\begin{Proposition}\label{} - 
If the initial wave function $ \Psi^h_{0}(\mathbf{x}_1,\textbf{x}_2,..,\textbf{x}_N)$ is factorized in the form:
\begin{equation}\label{eq:solschrodinger2}
 \Psi^h_0(\textbf{x}_1, \textbf{x}_2,..., \textbf{x}_N) =\Psi^h_0(\textbf{x}_G) \varphi^h_0(\textbf{x}'_1,\textbf{x}'_2,...,\textbf{x}'_N).
\end{equation}
then $ \Psi^h(\textbf{x}_1, \textbf{x}_2,.., \textbf{x}_N, t)$, 
solution to (\ref{eq:schrodinger1d}),(\ref{eq:hamiltonschrodinger1}) and (\ref{eq:schrodinger2}),
is written as the product of an external wave function $ \psi^h(\textbf{x}_G,t)  $ and an internal function $\varphi^h(\textbf{x}'_1,\textbf{x}'_2,..,\textbf{x}'_N,t)$~:
\begin{equation}\label{eq:solschrodinger3b}
 \Psi^h(\textbf{x}_1, \textbf{x}_2,.., \textbf{x}_N,t) =\psi^h(\textbf{x}_G,t) \varphi^h(\textbf{x}'_1,\textbf{x}'_2,..,\textbf{x}'_N,t)
\end{equation} 
where $\Psi^h(\textbf{x}_G,t) $ is the solution to Schrödinger's external equations
\begin{equation}\label{eq:schrodinger1}
i\hslash \frac{\partial \Psi^h(\textbf{x}_G,t) }{\partial t}=- \frac{\hbar ^{2}}{2M}
\triangle_{\textbf{x}_G} \Psi^h(\textbf{x}_G,t)+ M \textbf{g}. \textbf{x}_G \Psi^h(\textbf{x}_G,t)
\end{equation}
with the initial condition
\begin{equation}\label{eq:solschrodinger4}
\Psi^h(\textbf{x}_G,0) =\Psi^h_0(\textbf{x}_G)
\end{equation}
and where $ \Phi^h(\textbf{x}'_1,\textbf{x}'_2,..,\textbf{x}'_N ,t)$ is the solution to Schrödinger's internal equations:
\begin{equation}\label{eq:schrodinger2b}
i\hslash \frac{\partial \varphi^h(\textbf{x}'_1,\textbf{x}'_2,..,\textbf{x}'_N,t) }{\partial t}=- \sum_i\frac{\hbar ^{2}}{2 m_i}
\Delta_{\textbf{x}'_i} \varphi^h(\textbf{x}'_1,\textbf{x}'_2,..,\textbf{x}'_N,t)+ \sum_{i,j}U_{ij}(\mathbf{x}'_i -\textbf{x}'_j) \varphi^h(\textbf{x}'_1,\textbf{x}'_2,..,\textbf{x}'_N,t)
\end{equation}
with the initial condition
\begin{equation}\label{eq:solschrodinger5}
 \varphi^h(\textbf{x}'_1,\textbf{x}'_2,..,\textbf{x}'_N,0)=  \varphi^h_0(\textbf{x}'_1,\textbf{x}'_2,..,\textbf{x}'_N).
\end{equation}
\end{Proposition}

The proposal is obtained simply by replacing in the Schrödinger~(\ref{eq:schrodinger1d}), $\Psi^h(\textbf{x}_1, \textbf{x}_2,...., \textbf{x}_N,t)$ per $\psi^h(\textbf{x}_G,t) \varphi^h(\textbf{x}'_1,\textbf{x}'_2,...,\textbf{x}'_N,t)$. This yields~:
\begin{eqnarray*}
 \left(i\hslash \frac{\partial}{\partial t}- H\right) \Psi^h(\textbf{x}_1, \textbf{x}_2,.., \textbf{x}_N, t)=&&\psi^h(\textbf{x}_G,t) \left[ \left(i\hslash \frac{\partial}{\partial t}- H_{int}\right)\varphi^h(\textbf{x}'_1,\textbf{x}'_2,..,\textbf{x}'_N,t)\right]\\
&+&\varphi^h(\textbf{x}'_1,\textbf{x}'_2,..,\textbf{x}'_N,t)\left[ \left(i\hslash \frac{\partial }{\partial t} -H_{ext}\right) \psi^h(\textbf{x}_G,t)\right]=0.
\end{eqnarray*}
The decomposition of the total wave function as the product of an external wave function and an internal wave function is due to the independence of the external and internal variables thus allowing an exact decomposition of the Hamiltonian into its external and internal parts.

The fundamental property of the external wave function of a quantum system is that it can spread over time in space and be divided into several parts (see the interference experiments of section~\ref{sect:interpretationExterne}). 
It can be very large.
On the contrary, the internal wave function of a system remains confined to space, it does not spread and cannot divide without changing the nature of the system; this is what happens during ionization, nuclear fission or chemical reaction. 
The size of the internal wave function is the size commonly given for an atom or molecule and is often much smaller than the size of the external wave function (i.e. the width of the wave packet) as we will see in the examples in section~\ref{sect:interpretationExterne}.

When the quantum system is not composed of several particles but corresponds to an elementary particle such as a free electron,
it can also be associated with an internal wave function and an external wave function. 
Its external wave function is the wave function usually associated with an electron coming out of an electron gun of an electron microscope or a tungsten tip of a scanning tunnel microscope.
Although many physicists consider electrons to be point clouds, Schrödinger proposes to consider them as electronic clouds with a continuous charge distribution.
The internal wave function of a free electron is not known but could correspond to this Schrödinger electronic cloud.

\begin{Remarque}\textbf{- The role of gravity -} 
The independence between the external and internal variables of proposition 1 comes in part from the very special form of the linear gravitational potential $V_j(\textbf{x}_j) $ which depends linearly on the product $m_j \textbf{x}_j $, which gives us $ \Sigma_j V_j(\textbf{x}_j)= M \textbf{g} .\textbf{ x}_G $. Without this hypothesis, we would have had $ \Sigma_j V_j(\textbf{x}_j)= \Sigma_j V_j(x_G -\textbf{x}'_j)$ and an interaction between external and internal variables; and no exact breakdown!   In addition, with the linear potential $V_j(\textbf{x}_j)= m_j \textbf{g} .\textbf{x}_j$, gravity is transferred exactly to the external wave function. 
\end{Remarque}

In the general case of a quantum system with $N$ particles, there is interaction between the external and internal variables due to the role of the environment, and the external equations of Schrödinger (\ref{eq:schrodinger1}) and (\ref{eq:solschrodinger4}) and internal variables of Schrödinger (\ref{eq:schrodinger2b}) and (\ref{eq:solschrodinger5}) are only approximated. 
This is particularly the case when it is assumed that each particle $i$ admits a charge $q_i$ and is also subjected to an electrical potential $q_i V_q(\textbf{x}_i)$ that varies little on the scale of the quantum system: $V_q(\textbf{x}_i)\sim V_q(\textbf{x}_G)$. Under this assumption, the external field applying to the external wave function is written approximately:
\begin{equation}\label{eq:solspot}
V(\textbf{x}_G)= M \textbf{g} .\textbf{x}_G + \sum_i q_i V_q(\textbf{x}_i) \simeq   M \textbf{g} . \textbf{x}_G + Q V_q(\textbf{x}_G)
\end{equation}
with $Q=\Sigma_i q_i$. We then consider a generalization of Schrödinger's external equation (\ref{eq:schrodinger1}) by replacing $M \textbf{g}\cdot \textbf{x}_G$ by $V(\textbf{x}_G)$. The solution obtained will no longer be accurate, but will be a good approximation if the quantum system is not too shaken and remains stable during its evolution. This will no longer be the case if it disintegrates.
The case where there is an external magnetic field is taken into account in section~\ref{sect:interpretationExterne} in the Stern and Gerlach and EPR-B experiments. We can certainly generalize the external wave function to mesoscopic and macroscopic quantum systems that are neither atoms nor molecules.

\begin{Remarque}\textbf{- the $N$ individual functions of a N-body system -}
\label{rmq:emergence}
In addition to the external and internal wave functions associated with a $N$ particle system, each of the $N$ individual particles in the system must also be associated with an individual wave function. 
However, since the N particles interact, the individual functions are not accessible because they are intertwined with the other individual functions. 
The external (respectively internal) function of the complete system emerges from the entanglement of the $N$ individual functions.
For example, let us consider a proton and an electron both free in an empty and confined space; they each have an individual internal and external wave function. 
If the experimental conditions cause the proton to capture the electron, the hydrogen atom thus formed will have new internal and external wave functions that emerge from the individual internal and external wave functions.
An example of the emergence of the external wave function from two individual external wave functions is given in section~\ref{sect:EPR-B} for the EPR-B experiment.
\end{Remarque}

\subsection{Preparation of an internal wave function or an external wave function depending on the experiments}

In many studies, knowledge of the quantum system does not correspond to a total wave function but only to an external wave function or an internal wave function.
In particular, only the external wave function is considered for all measurement problems that are mostly related to the measurement of the position of the centre of mass. These are particularly the cases of atomic interferometry, spin measurement and energy.

The internal wave function will explain the values related to the energy spectrum and quantum jumps, but they will be measured by position measurements via an external wave function.
Thus in the debates of the 1927 Solvay Congress, de Broglie, with \emph{``the pilot wave''}, and Born, with \emph{`the statistical wave''}, implicitly considered the external wave function, while Schrödinger, with his \emph{``soliton''}, and Heisenberg, with his quantum jumps, implicitly considered the internal wave function.

\subsection{Classical external and internal analogies}

We have just seen that there is not always a simple relationship between the total wave function of a quantum system and the external and internal wave functions. The case of a simple product as in Proposition 1 is exceptional. In general, external variables influence internal variables and vice versa. The following two non-quantum analogies are instructive when it comes to understanding the types of interaction between the external and internal wave functions: the analogy with Couder's\emph{walkers} and the analogy with the solar system.

\subsubsection{Analogy with Couder's "walkers"}

The wave-particle duality seems to be a characteristic of the quantum world, having no equivalent in classical physics. Yves Couder and his team~\cite{Couder2005a, Couder2005b} 
have shown \textit{``that a drop bouncing off a vertically vibrating liquid surface can become self-propelled by its interaction with the surface wave it excites''}. The drop couples to the surface waves that its rebounds generate and spontaneously moves on the surface. The resulting object, called a "walker", combines the drop and its surface waves, possesing a dual nature that enables it to perform many of the classical quantum mechanics experiments: Young's slits~\cite{Couder2006}, tunnel effect~\cite{Eddi2009}, quantized orbits~\cite{Fort2010}.

Even if the walkers studied by Couder and his team present fundamental differences with the quantum case (system maintained by vibration, no Planck constant, existence of waves on a material medium), they show us that the wave-particle duality exists at the macroscopic level. The deformations of the drop during bounces constitute its internal evolution. The external evolution of the drop,
i.e. its center of mass, is governed by the surface wave it creates during each bounce and the liquid
vibrating exterior.

\subsubsection{Analogy with the solar system}
To study the evolution of a planet in the solar system, we 
breaks down the problem into two sub-problems: 1) the evolution of the center of mass of this planet (external evolution) for which the planet is reduced to a point, its center of mass. 2) the internal evolution of the planet in the reference frame of its center of mass; this evolution takes into account the fact that the planet is not a point but has a gaseous and/or solid structure and possibly a rotational movement on itself.
These two developments can be treated separately in the first 
approximation.
But depending on the experimental conditions of each planet, we can 
observe significant effects of external evolution on the internal evolution; for example, terrestrial tides due to very strong sunlight slightly deforming the solid structure of the Earth.

For a quantum system, we observe the same decomposition, an 
external evolution and internal evolution. We therefore have two functions: an external one that describes the evolution of the center of mass and an internal one defined in the reference frame of the center of mass.

However, depending on the experimental conditions, we will focus only on external variables or only on internal variables. This was the methodology we applied in the  double-preparation theory~\cite{Gondran2014c}. In many experiments, it is indeed possible to separate these two wave functions. This is the case, for example, of free particles such as 
in the Young or Stern and Gerlach or EPR-B experiments where only 
the external evolution of particles (i.e. the evolution of the particle center of mass) 
is studied, internal evolution being neglected. On the other hand, if we 
study the emission lines of a gas, only the internal evolution of 
gas particles is studied, external evolution being neglected.

\section{Interpretation of the external wave function}
\label{sect:interpretationExterne}

To interpret the external wave function, we will study 
how it evolves when the Planck constant is mathematically tended towards zero.

We consider the semi-classical variable change $\Psi^h(\textbf{x}_G,t)=\sqrt{\rho^{\hbar}(\mathbf{x}_G,t)} \exp(i
\frac{S^{\hbar}(\textbf{x}_G,t)}{\hbar})$, the density
$\rho^{\hbar}(\mathbf{x}_G,t)$ and the action $S^{\hbar}(\textbf{x}_G,t)$
being functions that depend a priori on $\hbar$. 

Schrödinger's external equations (\ref{eq:schrodinger1} and \ref{eq:solschrodinger4}) are decomposed by giving the Madelung equations~\cite{Madelung1926} (1926) which correspond to both
coupled equations:
\begin{equation}\label{eq:Madelung1}
\frac{\partial S^{\hbar}(\mathbf{x}_G,t)}{\partial t}+\frac{1}{2M}
(\nabla S^{\hbar}(\mathbf{x}_G,t))^2 +
V(\mathbf{x}_G)-\frac{\hbar^2}{2M}\frac{\triangle
\sqrt{\rho^{\hbar}(\mathbf{x}_G,t)}}{\sqrt{\rho^{\hbar}(\mathbf{x}_G,t)}}=0
\end{equation}
\begin{equation}\label{eq:Madelung2}
\frac{\partial \rho^{\hbar}(\mathbf{x}_G,t)}{\partial t}+
div(\rho^{\hbar}(\mathbf{x}_G,t)
\frac{\nabla S^{\hbar}(\mathbf{x}_G,t)}{m})=0  
\end{equation}
with the initial conditions
\begin{equation}\label{eq:Madelung3}
\rho^{\hbar}(\mathbf{x}_G,0)=\rho^{\hbar}_{0}(\mathbf{x}_G) \qquad and
\qquad S^{\hbar}(\mathbf{x}_G,0)=S^{\hbar}_{0}(\mathbf{x}_G). 
\end{equation}
Here, $V(\textbf{x}_G)$ is a general potential like (\ref{eq:solspot}). In this section we study the convergence of the density $\rho^{\hbar}(\mathbf{x}_G,t)$ and of the action
$S^{\hbar}(\textbf{x}_G,t)$ of the Madelung equations when the Planck constant $\hbar$ is mathematically tended towards 0.

We will limit ourselves to "prepared as non-discerned" quantum systems, i.e. such that the initial probability density $\rho^{\hbar}_{0}(\mathbf{x}_G)$ and the initial action $S^{\hbar}_{0}(\mathbf{x}_G)$ of the external wave function are functions $\rho_{0}(\mathbf{x}_G)$ and $ S_{0}(\mathbf{x}_G)$ independent of    $\hbar$. This is the case of a set of particles without interaction between them and prepared in the same way: jets of free particles or in a gravity field as in the Shimizu~\cite{Shimizu1992} experiment with cold atoms, or jets of fullerenes in a Young's slits experiment.

\begin{Theoreme}\label{r-th1}- When $\hbar$ tends towards 0, density
$\rho^{\hbar}(\textbf{x}_G,t)$ and action $S^{\hbar}(\textbf{x}_G,t)$, solutions to Madelung equations (\ref{eq:Madelung1}-\ref{eq:Madelung3}), converge to $\rho(\textbf{x}_G,t)$ and  $S(\textbf{x}_G,t)$,
solutions of statistical Hamilton-Jacobi equations:
\begin{equation}\label{eq:statHJ1b}
\frac{\partial S\left(\textbf{x}_G,t\right) }{\partial
t}+\frac{1}{2m}(\nabla S(\textbf{x}_G,t) )^{2}+V(\textbf{x}_G)=0
\end{equation}
\begin{equation}\label{eq:statHJ2b}
S(\textbf{x}_G,0)=S_{0}(\textbf{x}_G)
\end{equation}
\begin{equation}\label{eq:statHJ3b}
\frac{\partial \mathcal{\rho }\left(\textbf{x}_G,t\right) }{\partial
t}+ div \left( \rho \left( \textbf{x}_G,t\right) \frac{\nabla
S\left( \textbf{x}_G,t\right) }{m}\right) =0
\end{equation}
\begin{equation}\label{eq:statHJ4b}
\rho(\mathbf{x}_G,0)=\rho_{0}(\mathbf{x}_G).
\end{equation}
\end{Theoreme}
The demonstration was made in~\cite{Gondran2014c} whose principle we recall briefly. In the case of indiscriminate prepared quantum systems, the external wave function $\Psi(\textbf{x}_G,t) $ at the time $t$ is derived from the initial external wave function by the Feynman path integral~\cite{Feynman1965}~: $ 
\exp\left(\frac{i}{\hbar}S_{cl}(\textbf{x}_G,t;\textbf{x}_{0})\right)
\Psi_{0}(\textbf{x}_{0})d\textbf{x}_0= F(t,\hbar) \int
\sqrt{\rho_0(\textbf{x}_0)}\exp\left(\frac{i}{\hbar}( S_0(\textbf{x}_0) + S_{cl}(\textbf{x}_G,t;\textbf{x}_{0})\right)
d\textbf{x}_0$.
The stationary phase theorem then shows that, if $\hbar$ tends towards $0$, we have $\Psi(\textbf{x}_G,t)\sim\exp\left(\frac{i}{\hbar}min_{\textbf{x}_0}( S_0(\textbf{x}_0)+
S_{cl}(\textbf{x}_G,t;\textbf{x}_{0})\right)$, which implies that the external quantum action $S^{\hbar}(\textbf{x}_G,t) $ converges to the function~:
\begin{equation}\label{eq:solHJminplus}
S(\textbf{x}_G,t)=min_{\textbf{x}_0}( S_0(\textbf{x}_0)+
S_{cl}(\textbf{x}_G,t;\textbf{x}_{0})
\end{equation}
which is the solution of the Hamilton-Jacobi equation (\ref{eq:statHJ1b}) with the initial condition (\ref{eq:statHJ2b}). 
The equation (\ref{eq:solHJminplus}) is an integral in the Minplus analysis which corresponds to a new branch of mathematics (cf. Maslov~\cite{Maslov1992}, Gondran~\cite{Gondran1995a, Gondran2008a} chapter 7) which studies certain non-linear problems by a linear approach. It is constructed by replacing the conventional scalar product $(f,g) = \int_{X}f(x)g(x) dx$ by the Minplus scalar product $(f,g)_{min+} =\,\underset{x\in X}{\inf }\left\{ f(x)+g(x) \right\}$. This equation (\ref{eq:solHJminplus}) is therefore the analog in classical mechanics of the Feynman path integral and we called it~\cite{Gondran2014c} \textbf{the Minplus path integral}.

The demonstration continues by noting that, like quantum density $\rho^{h}(\textbf{x}_G,t)$
satisfies the continuity equation (\ref{eq:Madelung2}), we deduce, since $S^{h}(\textbf{x}_G,t)$ tends towards $S(\textbf{x}_G,t)$, that
$\rho^{h}(\textbf{x}_G,t)$ converges to the classical density
$\rho(\textbf{x}_G,t)$, which satisfies the continuity equation (\ref{eq:statHJ3b}).

Thus, if the external wave function is prepared as non-discerned, the Madelung equations of this external wave function converge to the Hamilton-Jacobi statistical equations. These statistical Hamilton-Jacobi equations correspond to a set of classical particles, without interaction
between them and subjected to an external potential field $V(\mathbf{x})$, and of which we only know, at the initial moment,
the probability density $\rho _{0}\left( \mathbf{x}\right)$ and
the speed field $\mathbf{v_{0}(\textbf{x})}$ through the
of the initial action $S_0(\textbf{x})$ ($\mathbf{v_{0}}(\textbf{x})=
\frac{\nabla S_0(\textbf{x})}{m}$). These are prepared as non-discerned classical particles like prepared as non-discerned quantum particles. For these prepared as non-discerned classical particles, the velocity of the center of mass of the classical particle is given in each point $ \left(
\mathbf{x,}t\right)$ by:
\begin{equation}\label{eq:eqvitesse1}
\mathbf{v}\left( \mathbf{x,}t\right) =\frac{\mathbf{\nabla}S\left( \mathbf{%
x,}t\right) }{m}\text{\ \ \ }
\end{equation}

Equation (\ref{eq:eqvitesse1}) shows that the $S\left(
\mathbf{x,}t\right) $ of the Hamilton-Jacobi equations (\ref{eq:statHJ1b}) defines the velocity field at any point ($\textbf{x},t$) to
from the speed field $\frac{\nabla S_0(\textbf{x})}{m} $ to
the initial moment. So, if we
give itself the initial position $\textbf{x}_{init}$ of the center of mass of a prepared as non-discerned classical particle, we deduce by (\ref{eq:eqvitesse1}) the
trajectory of the particle's center of mass. Hamilton-Jacobi's action $S\left(
\mathbf{x,}t\right)$ is therefore a field that \emph{``pilots''} the movement
of the classical particle.

The indetermination on the position of the center of mass of a quantum system therefore corresponds to
the indetermination on the position of the center of mass of a conventional system of which
only the initial distribution density has been defined.
Like Hamilton-Jacobi's action for a prepared as non-discerned classical particle, the external wave function of a quantum system is not sufficient to define the position of the center of mass of the quantum system; it is necessary
 add its initial position and it is therefore natural to introduce
the Broglie-Bohm trajectories for the center of mass of a quantum system. Its speed at the moment $t$ is given by
\cite{deBroglie1927,Bohm1952}:
\begin{equation}\label{eq:quantum velocity}
\textbf{v}^{\hbar}(\textbf{x}_G,t) = \frac{1}{m} \nabla S^\hbar (\textbf{x}_G,t)
\end{equation}
which satisfies the continuity equation (\ref{eq:Madelung2}).

In the rest of this section, we give three experiments where only the external wave function is considered. 
Indeed, in many studies, knowledge of the internal wave function is not necessary and can therefore be neglected.
This is the case of particle jet without interactions between them that we find in very many experiments: diffraction, interference, spin measurement.  
This is particularly the case for all measurement problems, which are most often related to the measurement of the position of the center of mass. 
We first examine the case of atomic interferometry and then that of spin measurement, and finally the EPR-B experiment.

\subsection{Practice case 1~: interference of the external wave of the molecule $C_{60}$}
Young or Mach-Zehnder interferometry experiments are examples where only the external wave function of a particle interferes with itself.
The internal structure is not necessary to understand the experiment and is omitted in the calculations.

\begin{figure}[ht]
\includegraphics[width=0.8\linewidth]{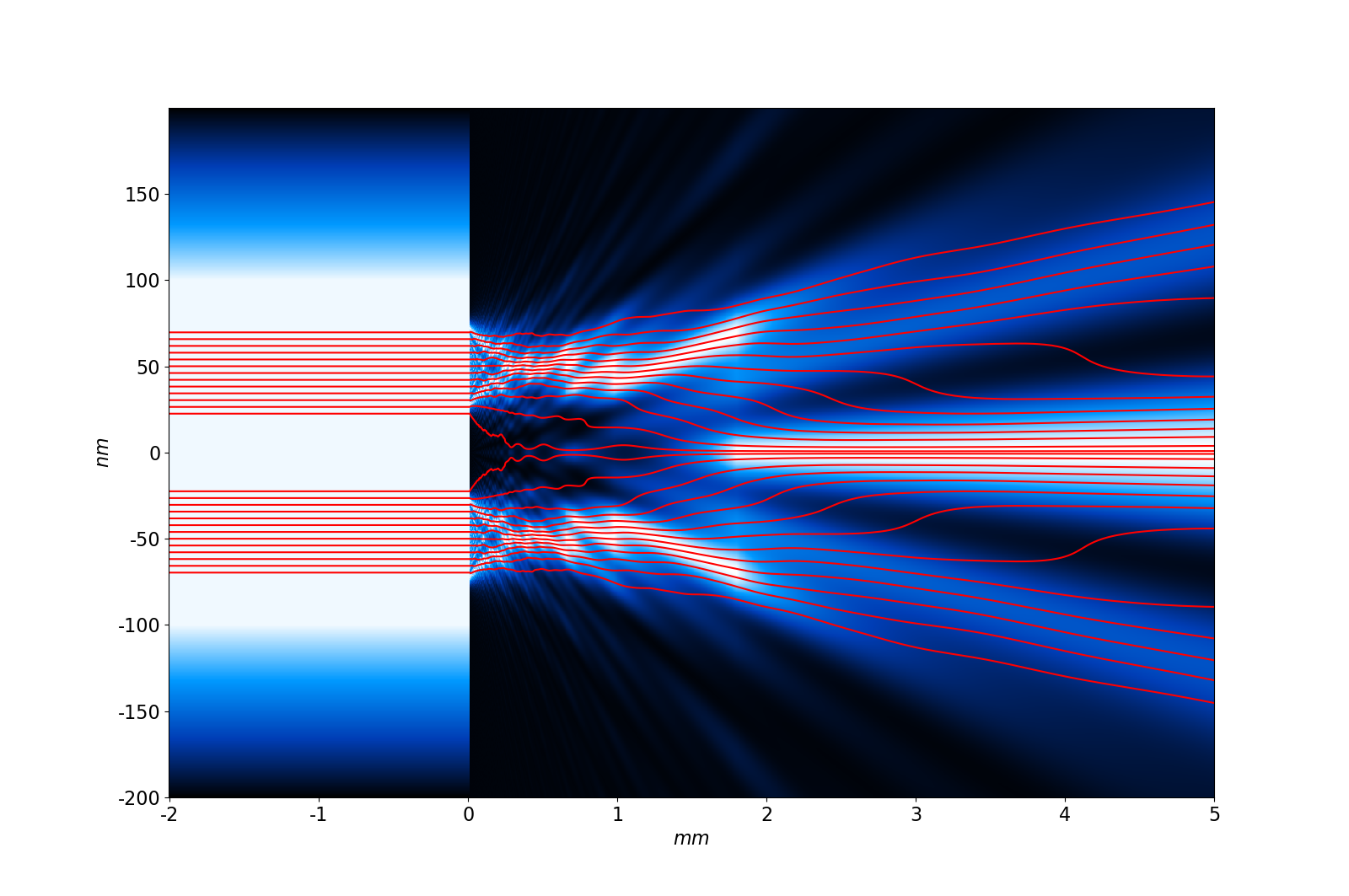}
\caption{\label{fig:traj-Young2} Simulation of 24 trajectories of $C_{60}$ under the experimental conditions of~\cite{Nairz2003} corresponding to 24 different starting points of the center of mass of a $C_{60}$.}
\end{figure}

The figures~\ref{fig:traj-Young1} and~\ref{fig:traj-Young2} represent a simulation of Young's slits experiment with fullerene molecules $C_{60}$
under the conditions of the experiment conducted by Nairz, Arndt and Zeilinger~\cite{Nairz2003}. The two slits are spaced $100~nm$ (center to center), each with a wisth of $55~nm$. The average speed of the molecules is $200~m/s$, which corresponds to a wavelength of $2.8~pm$. 
The standard of the external wave function is represented in the blue figures: the lighter the blue, the higher the density.

In figure~\ref{fig:traj-Young1}, the external wave function of a molecule of $C_{60}$ is represented at fifteen different times, from two millimeters before the slits (far left) to 5 millimeters after the slits (far right).
The red line corresponds to the trajectory of the center of mass of a molecule $C_{60}$ whose initial position (before the slits) was randomly drawn in the initial wave packet~\footnote{Videos from this experiment are available at the addresses~: \texttt{vimeo.com/350139153} and \texttt{vimeo.com/350132498}}.

The internal wave function of a molecule $C_{60}$ that defines the internal structure is schematically represented on the figure by a drawing of the arbitrarily magnified molecule ($\times13$) because its size is only $1~nm$. 
The internal wave function remains unchanged throughout the experiment, its size remains $1~nm$ before and after the slits, and it does not interact with the external wave function that it transforms deeply.
This experiment can be considered crucial because it is difficult to imagine that the internal wave function of a molecule $C_{60}$ does not pass through only one of the slits. It is its external wave function that passes through both slits at the same time.
When measuring the impact of the molecule $C_{60}$ to $5~mm$ after the slits, it is the internal wave function that interacts with the detection screen and produces the impact.

\begin{figure}[ht]
\caption{\label{fig:traj-Young1} Simulation of the evolution of the external (blue) and internal (white, magnified 13 times) wave functions of a molecule of $C_{60}$ under the experimental conditions of~\cite{Nairz2003} at fifteen different times every $2.5~\mu s$ (i.e. every $0.5~mm$). The slits are placed at $0~mm$ and are spaced $100~nm$ and have a width of $55~nm$.}
\end{figure}

The figure~\ref{fig:traj-Young2} describes the same experiment, the density of the external wave function is continuously represented from $2~mm$ before the slots to $5~mm$ after. 24 trajectories of the center of mass of a molecule of $C_{60}$ are represented by a red line corresponding to 24 different starting points of the center of mass.

Thus for a \textit{\textbf{" prepared as non-discerned" }} external wave function, $\rho_0(\textbf{x}_G) $ corresponds to the initial probability density of the center of mass and $\rho^\hbar(\textbf{x}_G,t) $ to the probability density of the center of mass at the time t.
The \emph{prepared as non-discerned} external wave function can therefore be considered as a probability wave to which Born's statistical interpretation applies. We also find the point of view of Dirac who, in 1930 writes~\cite{Dirac1930}:
 \textit{``In quantum mechanics, particles are connected to waves that direct them and give rise, under appropriate conditions, to phenomena of interference and diffraction''}, and he adds that it is only a question of \textit{"\textit{one and the same reality}"}.

\begin{Remarque}\textbf{- Convergence from $\hbar$ to $0$ -} 
In the theorem~\ref{r-th1}, we assume that $\hbar$ tends towards $0$. 
However, physically $\hbar$ is never equal to $0$ and cannot tend towards $0$ since it is a constant. and more generally we never have a trajectory that can be called classic. 
All trajectories are quantum and the so-called classical trajectories are approximations of quantum trajectories for which the term in $\hbar$ is negligible compared to the other terms. 
A very simple example is that of a quantum object defined by a Gaussian wave packet of center $(x_0, y_0, z_0)$ and standard deviation $(\sigma_{0x}, \sigma_{0y}, \sigma_{0y}, \sigma_{0z})$ and whose center of mass 
 at the initial position
$(x_G(0),y_G(0),z_G(0))$ and the initial speed
$\textbf{v}_{0}=(v_{0x},v_{0y},v_{0z})$. If $\textbf{g}=(0.0,g)$ is the only external force, then the particle's center of mass satisfies the following Broglie-Bohm trajectory:
\begin{eqnarray}
X^h(t)&=&x_G(0)+v_{0x}t+(x_0-x_G(0))\left(1-\frac{\sigma_{\hbar x}(t)}{\sigma_{0x}}\right)
\\
Y^h(t)&=&y_G(0)+v_{0y}t+(y_0-y_G(0))\left(1-\frac{\sigma_{\hbar y}(t)}{\sigma_{0y}}\right)
\\
Z^h(t)&=&z_G(0)+v_{0z}t-\frac{g t^{2}}{2m}+(z_0-z_G(0))\left(1-\frac{\sigma_{\hbar z}(t)}{\sigma _{0z}}\right)
\end{eqnarray}
with $\sigma_{\hbar i}(t)=\sigma_{0i}\sqrt{1+\left(\frac{\hbar t}{2m\sigma_{0i}^2}\right)^2}$ and $i=x,y,z$. We verify that this de Broglie-Bohm trajectory tends towards the trajectory we call classic and the same starting point when we converge $\hbar$ towards $0$. In reality, it is not $\hbar$ that tends towards $0$, but the last term of the three equations that is negligible compared to the other terms if we consider a classical object as a stone, whereas if the object is an electron, an atom or a molecule, this term is not necessarily negligible. We observe that if, $m$, the mass of the object increases, then the situation is the same as if $\hbar$ decreases. It is these trajectories that we used to simulate Shimizu's experiment~\cite{Shimizu1992} on Young's slits with cold atoms; we thus showed~\cite{Gondran2005a} why such atoms with different wave functions were coherent enough to interfere.
\end{Remarque}

Thus, the external wave function corresponds, within the semi-classical limit, to the evolution of the center of mass of quantum systems. In relation to the scale of de Broglie wavelength, we return to the classical world.

\subsection{Case study 2~: spin measurement by the external wave for the Stern and Gerlach experiment}

Let us study the practical case of the spin which is a property also carried by the external wave function of the particle.
Indeed, the measurement of the spin of a particle is a measure of the position of its center of mass.
Let's consider the Stern and Gerlach experiment in measuring the spin of a silver atom. These atoms have, at the output of the speaker of $\mathbf{E}$, a velocity $\textbf{v}$ parallel to $(Oy)$. They cross an electromagnet $\mathbf{A}_{1}$ before
condensing on a plate $\mathbf{P}_{1}$ (Fig.~\ref{fig:schemaSG}).

The magnetic moments of these silver atoms have been prepared in a pure state $(\theta_0, \varphi_0)$~~\footnote{To prepare atoms all in the same state, the jet of atoms is first passed through a first Stern and Gerlach apparatus, and only one of the two outputs is kept that is judiciously oriented to obtain the desired pure state.}
the electromagnet $\mathbf{A}_{1}$ at the initial instant $t=0$ each atom can be described by the Gaussian spinor in $x$
and $z$~:
\begin{equation}\label{eq:psi-0}
    \Psi^{0}(x,z) = (2\pi\sigma_{0}^{2})^{-\frac{1}{2}}
                      e^{-\frac{(z^2+x^2)}{4\sigma_0^2}}
                      \binom{\cos \frac{\theta_0}{2}e^{ i\frac{\varphi_0}{2}}}
                            {i\sin\frac{\theta_0}{2}e^{-i\frac{\varphi_0}{2}}}
\end{equation}
with $\textbf{r}=(x,z)$. The variable $y$ is processed in the classic way with $y= vt$.
For the silver atom~\cite{Cohen1977}, we have $m = 1.8\times 10^{-25}$ kg, $v =
500$\ m/s, $\sigma_0$=10$^{-4}$m. In the
(\ref{eq:psi-0}), $\theta_0$ and $\varphi_{0}$ are the angles
polarities characterizing the initial orientation of the vector
representing the magnetic moment, $\theta_0$ corresponds to the angle
with $(Oz)$. Here, we have a pure state and all silver atoms have the same orientation of the magnetic moment. 

\begin{figure}
\centering
\caption{\label{fig:schemaSG} Diagram of the Stern and Gerlach experiment: a stream of silver atoms, prepared in a pure state $(\varphi_0,\theta_0)$ and coming from the enclosure $\mathbf{E}$ passes through an inhomogeneous magnetic field (magnet $\mathbf{A_1}$), then separates into two distinct beams giving on the plate $\mathbf{P_1}$ two distinct spots of intensity $N^+$ and $N^-$.}
\end{figure}
Most quantum mechanics textbooks do not take into account the spatial extension $f(\textbf{r})=(2\pi\sigma_{0}^{2})^{-\frac{1}{2}}
                      e^{-\frac{(z^2+x^2)}{4\sigma_0^2}} $ of the spinor
(\ref{eq:psi-0}) and simply take the wave function
in Hilbert's space of dimension 2 generated by $|0\rangle= \binom
{1}{0}$
and $|1\rangle= \binom {0}{1} $:
\begin{equation}\label{eq:psi-copenhagen}
|\psi^0\rangle=\cos \frac{\theta_0}{2} and^{i \frac{\varphi_0}{2}}
|0\rangle + i \sin \frac{\theta_0}{2} e^{-i
\frac{\varphi_0}{2}}|1\rangle
\end{equation}
By not retaining for the
quantum system only the wave function (\ref{eq:psi-copenhagen})
with orientation ($\theta_0$, $\varphi_0$), you lose a part
of the spinor (\ref{eq:psi-0}) and we only keep the
statistical character. The spatial extension of the spinor (\ref{eq:psi-0})
takes into account the initial position ($x_0$,
$z_0$) of the particle's center of mass (external variable) and makes the system quantum's evolution
(wave function + position) deterministic.

The initial spinor~(\ref{eq:psi-0}) is actually only the external wave function of the silver atom. 
The internal wave function of the silver atom is not useful to describe the experiment. 
However, the fact that the silver atom has a spin comes from the addition of all the spin moments of its internal electrons. Atoms with an odd number of electrons, such as the silver atom, which has 47 electrons, have a half spin.
The size of the internal wave function is about twice that of the atomic radius of the silver atom, or about $30~nm$. 
while the initial size of the external wave function, i.e. the width of the initial wave packet, is $3\times\sigma_0=3\times10^5~nm$,
which is 4 orders of magnitude larger than the internal one.

The evolution of spinor $\Psi=\binom{\psi_{+}}  {\psi_{-}}$
                            in a magnetic field
                            $\textbf{B}=(
                            B_x,B_y,B_z)$ is then given by the
Pauli equation~\cite{Cohen1977}:
\begin{equation}\label{eq:Pauli}
    i\hbar \left( \begin{array}{c} \frac{\partial \psi _{+}}{\partial t}
                                   \\
                                   \frac{\partial \psi _{-}}{\partial t}
                  \end{array}
           \right)
    =-\frac{\hbar ^{2}}{2m}\Delta
                           \left( \begin{array}{c} \psi _{+}
                                                   \\
                                                   \psi _{-}
                                  \end{array}
                           \right)
     +\mu _{B}\textbf{B}\boldsymbol\sigma \left( \begin{array}{c} \psi _{+}
                                                                  \\
                                                                  \psi _{-}
                                                 \end{array}
                                          \right)
\end{equation}
where $\mu_B=\frac{e\hbar}{2m_e}$ is Bohr's magneton,
$\boldsymbol\sigma=(\sigma_{x},\sigma_{y},\sigma_{z})$ corresponds
to Pauli's three matrices and where $\textbf{B}\boldsymbol\sigma$ corresponds to the $ B_x \sigma_{x}+ 
B_y \sigma_{y}+B_z \sigma_{z}$.

Silver atoms pass throught an electromagnetic field
$\textbf{B}$ oriented mainly along the $(Oz)$ axis,
$B_{x}=B'_{0} x$; $B_{y}=0$; $B_{z}=B_{0} -B'_{0} z$, with
$B_{0}=5$ Tesla, $B'_{0}=\left| \frac{\partial B}{\partial
z}\right| =- \left| \frac{\partial B}{\partial x}\right|= 10^3$
Tesla/m over a length $\Delta l=1~cm$. Upon exiting of the magnetic field, the particles are
free up to the plate $P_1$ placed at a distance $D=20~cm$.
The particle passes the time $\Delta t=\frac{\Delta l}{v}= 2\times
10^{-5} s$ in the magnetic field. At the exit of this field, we show~\cite{Gondran2005b,Gondran2016a} that at the moment $t+ \Delta t$ $(t
\geq 0)$, the external spinor is equal to:

\begin{equation}\label{eq:fonctionapreschampmagnétique}
\Psi (x_G,z_G,t+\Delta t) = \left(
\begin{array}{c}
                                R_{+}e^{i\frac{S_{+}}{\hbar }} \\
                                R_{-}e^{i\frac{S_{-}}{\hbar }}
                            \end{array}
                     \right)
\simeq\left(
\begin{array}{c}
                                \cos \frac{\theta_0}{2}
                 (2\pi\sigma_0^2)^{-\frac{1}{2}}
                 e^{-\frac{(z_G-z_{\Delta}- ut)^2 + x_G^2}
                 {4\sigma_0^2}} e^{i\frac{m u z_G + \hbar \varphi_+}{\hbar }} \\
                                i \sin \frac{\theta_0}{2}
                (2\pi\sigma_0^2)^{-\frac{1}{2}}
                e^{-\frac{(z_G +z_{\Delta}+ ut)^2 + x_G^2}
                {4\sigma_0^2}} e^{i\frac{-
    muz_G + \hbar \varphi_- }{\hbar }}
                            \end{array}
                     \right)
\end{equation}
with~:
\begin{equation} \label{eq:zdeltavitesse}
    z_{\Delta}=\frac{\mu_B B'_{0}(\Delta
    t)^{2}}{2 m}=10^{-5}m,~~~~~~~u =\frac{\mu_B B'_{0}(\Delta t)}{m}=1 m/s.
\end{equation}

In the Broglie-Bohm interpretation, the external spinor will define the trajectory $X_G(t)= (x_G(t),z_G(t))$ of the silver atom's center of mass from its initial position $X_G(0)=(x_G(0), z_G(0)$ by the formula~\cite{Bohm1955a,Takabayasi1955}: 
\begin{equation}\label{eq:vitesse}
\dfrac{dX_G(t)}{dt}=\frac{\hbar}{2m \rho} Im{(\Psi^\dag\boldsymbol\nabla \Psi)}\vert_{\textbf{x}=X_G(t)}
\end{equation}
where $\Psi^\dag=(\Psi^{+*}, \Psi^{-*})$ and $\rho=\Psi^\dag\Psi$.  
Bohm et al.~\cite{Bohm1955a} define a spin vector field $\mathbf{s}$ as~:
\begin{equation}\label{eq:spinvector}
\mathbf{s}(\mathbf{x},t)= \frac{\hbar}{2\rho}\Psi^\dag(\textbf{x},t)\sigma\Psi(\textbf{x},t)=\frac{\hbar}{2}(sin\theta~ sin\varphi, sin\theta ~cos\varphi, cos\theta).
\end{equation}	
The spin vector of an individual particle is evaluated along its trajectory as follows: $\mathbf{s}= \mathbf{s}(X_G(t),t)$. 
This spin vector is totally defined by the spinor and the position of the particle's center of mass.

\begin{figure}[ht!]
 \begin{center}
\includegraphics[width=0.70\textwidth]{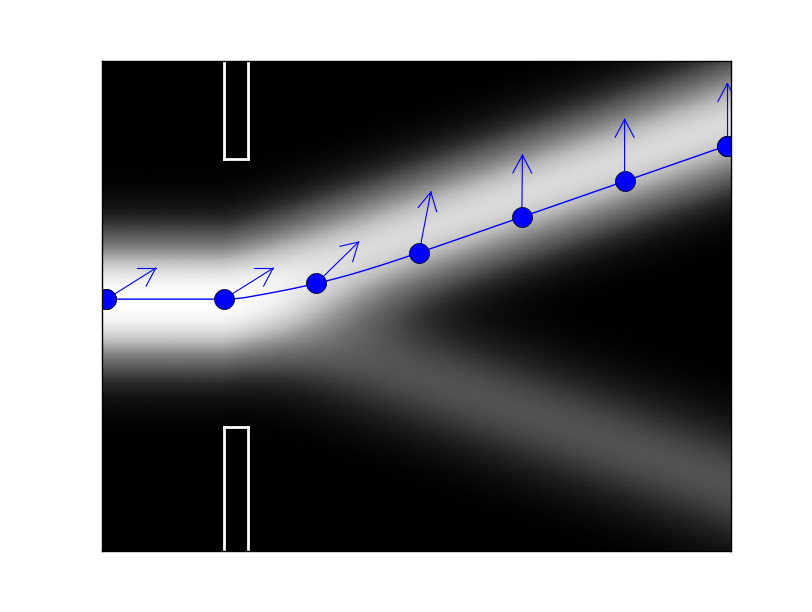}
 \label{fig:etat_pur_b}
\caption{ The arrows indicate the $\theta$ orientation of the spin vector $\mathbf{s}$ (initially $\theta_0=\pi/3$) along the path. The position of the particle exists before the measurement;
the particle then follows a deterministic trajectory and the impact on the screen only reveals its position.}
\label{fig:etat_pur}
 \end{center}
\end{figure}

Figure~\ref{fig:etat_pur} represents $\rho(z_G,t)=\int\Psi^\dag(x_G,z_G,t)\Psi(x_G,z_G,t)dx_G$, the probability density of presence of the silver atom for the values $\theta_0=\pi/3$ and $\phi_0=0$.
The axis $(Oy)$, of the jet propagation, is on the abscissa ($y=v_y t$) and the axis $(Oz)$ on the ordinate (the variable $x$ is not represented because the wave remains Gaussian along this axis). 
The magnet $A_1$ is represented on the figure by the two white rectangles, it is $\Delta l=1cm$ long and there is $D=20cm$ of free travel before the atom measurement on the $P_1$ detection screen.
A trajectory is also represented in figure~\ref{fig:etat_pur} with its spin $\mathbf{s}(X_G(t),t)$ along this trajectory. If the position of the particle's center of mass is located at the top of the wave packet, as shown in the figure, then the particle will be \emph{measured} in spin UP; if the initial position is lower, it will be measured DOWN.

As particles where the spin is not involved, the external spinor, which only uses the resolution of the Pauli equation on the external variables, gives the same statistical results as the usual quantum mechanics for the Stern-Gerlach experiment. Its resolution enables us~\cite{Gondran2016a} to propose a clear interpretation of the spin measurement by demonstrating the assumptions of the measurement and the reduction of the wave packet. The Stern and Gerlach experiment is not the measure of spin projection along the $(Oz)$ axis,
but the straightening of the spin orientation either in the direction of the magnetic field gradient or in the opposite direction.
The result depends on the initial position of the particle's center of mass in the external wave function. 
It is a simple explanation of the non-contextuality of measuring spin along different axes. 
The duration of the measurement is the time required for the particle to straighten its spin in the final direction.
The value \emph{"measured"} (the spin) is not a pre-existing value like the mass and charge of the particle but a contextual value in accordance with the Kochen and Specker theorem.

\subsection{Case study 3: simulation of the particles involved in the EPR-B experiment}
\label{sect:EPR-B}
For the EPR-B experiment, the total spinor (singulet state of the two massive entangled particles) is still only the external spinor. But as with the Stern and Gerlach experiment, the initial singulet wave function has a spatial extension:
 \begin{equation}\label{eq:singulet_aes}
     \Psi_{0}(\textbf{x}_G^A,\textbf{x}_G^B) =\frac{1}{\sqrt{2}}f(\textbf{x}_G^A) f(\textbf{x}_G^B)(|+_{A}\rangle |-_{B}\rangle - |-_{A}\rangle | +_{B}\rangle)
 \end{equation}
where $\textbf{x}_G^A$ and $\textbf{x}_G^B$ 
are the centres of gravity of the particles A and B and where
 $f(\textbf{x})=(2\pi\sigma_{0}^{2})^{-\frac{1}{2}}e^{-\frac{\textbf{x}^2}{4\sigma_0^2}}$.

It is possible to find this singulet spinor~(\ref{eq:singulet_aes}) from the Pauli principle.
To do so, we assume that at the moment of the creation of the two entangled particles $A$ and $B$, each of the particles has an initial wave function $\Psi_0^A(\textbf{x}_G^A, \theta^A_0,
 \varphi^A_0)$ and $\Psi_0^B(\textbf{x}_G^B, \theta^B_0, \varphi^B_0)$ of type~(\ref{eq:psi-0}) with opposite spins~: $\theta_0^B= \pi-\theta_0^A$, $\varphi_0^B= \varphi_0^A -\pi$.

If then we apply the Pauli principle, which stipulates that the entangled two-body must be antisymmetric, it is written:
\begin{eqnarray}\nonumber
 \Psi_0(\textbf{r}_A,\theta_A, \varphi_A,\textbf{r}_B,\theta_B, \varphi_B)&=&
 \Psi^0_A(\textbf{r}_A,\theta_A, \varphi_A)\Psi^0_B(\textbf{r}_B,\theta_B,
  \varphi_B)-\Psi^0_A(\textbf{r}_B,\theta_B, \varphi_B)\Psi^0_B(\textbf{r}_A,\theta_A, \varphi_A)\\
&=&- e^{i \varphi_A} f(\textbf{r}_A)f(\textbf{r}_B)(|+_{A}\rangle
|-_{B}\rangle - |-_{A}\rangle|++_{B}\rangle)\label{eq:singuletEPR}
\end{eqnarray}
which is the singulet state with spatial extension~(\ref{eq:singulet_aes}), factor-wise. Again this spatial extension is essential to correctly solve the Pauli equation in space because it is necessary to take into account the position of the atom in its external wave function. 

Then we measure the spin of the two particles one after the other. 
We show mathematically~\cite{Gondran2016a} that the first measured particle, particle $A$, behaves in the first Stern-Gerlach apparatus in the same way as if it were not entangled.

During the \emph{measure} of A, the density of the particle $B$ also evolves as if it were not entangled~\cite{Gondran2016a}. 
These two properties can be experimentally tested as soon as the EPR-B experiment with atoms is feasible.
During the \emph{measure} of A, the spin of the particle $B$ straightens up to always be in opposition to the spin of the particle $A$~\cite{Gondran2016a}.
The second measure is a Stern-Gerlach measure with specific orientations. 
We then find perfectly the results of quantum mechanics and the violation of Bell's inequalities.

As with the Stern and Gerlach experiment, the external spinor of the entangled state, which uses only the resolution of Pauli's equation on the external variables of the two particles yields, for the EPR-B, the same statistical results as the usual quantum mechanics. 
Quantum particles have a local position like a conventional particle, but also have a non-local behaviour due to the entangled external wave function.

We can refer to our article \emph{``Replacing the Singlet Spinor of the EPR-B Experiment in the Configuration Space with two Single-particle Spinors in Physical Space''}~\cite{Gondran2016a} where we show precisely how these three external spinors interfere, the singulet spinor with spatial extension which verifies the Pauli equation and the spinors of the two particles entangled with their spatial extensions.

\section{Schrödinger's interpretation of the internal wave function: quantum chemistry}
\label{sect:interpretationInterne}
For the internal wave function, the interpretation is more open~: should we take the interpretation of dBB~? or the Copenhagen interpretation restricted to the internal wave function~? or the interpretation proposed since 1926 by Schrödinger. It is with it, which is little known, that we will begin; but at the end of the section, we will discuss the other two.

Schrödinger proposed in 1926, then at the Solvay congress in 1927 and finally in 1952~\cite{Schrodinger1952} during a strong controversy with the Göttingen-Copenhagen school, an interpretation of the wave function that was both realistic and deterministic, as reported by Born in his 1954 Nobel lecture~\cite{Born1955}:  \textit{``Schrödinger thought that his wave theory
made it possible to return to deterministic
classical physics. He proposed (and he has
recently~\cite{Schrodinger1952} emphasized his proposal a new)
to dispensewith the particle reprentation entirely,
and instead of speaking of electrons as particles, to consider them as
continuous density distributions $|psi|^2 $, or electrical density $e|\psi|^2 $.''}
One can understand the criticisms of such an interpretation for the total wave function, but the arguments against it are no longer valid a priori for the internal wave function. The most important criticism corresponded to the contradiction between what Schrödinger considers to be the most important: \textit{``the particles are narrow wave packets''} and the fact that the external function spreads over time as Born points out in the rest of his Nobel speech.

The great difficulty in defining a reliable interpretation of the internal wave function of a quantum system of N particles is that the internal wave function is defined in the space of 3N dimension configurations and no explicit solutions are known.
However, we will use two specific examples of a single particle, the harmonic oscillator and the electron in the hydrogen atom as Schrödinger did, and then we will propose a generalization to the N-particle case.

\subsection{Schrödinger's interpretation of the internal wave function for a single particle}

The objective of Schrödinger's 1926 article, "\textit{The continuous transition from micro-to macro-mechanics}"~\cite{Schrodinger1926}, is to \textit{``demonstrate \textbf{in concreto} [for this chosen case of the harmonic oscillator] the transition to macroscopic mechanics by showing that a group of proper vibrations of high order-number n ("quantum number") and of relatively small order-number differences ("quantum number differences") may represent a "particle", which is executing the "motion", expected from the usual mechanics, i.e. oscillating with the frequency $\nu_0 $.''}

He considers the classic problem of the Hamiltonian harmonic oscillator in dimension 1 $ H=\frac{p_x^2}{2m}+ \frac{1}{2}m \omega^2x^2$. He then looked for the solution of the Schrödinger equation for a particular initial condition  that can be written today as
\begin{equation}\label{eq:coh0}
\Psi_0^h(x) = (2\pi \sigma_h)^{-\frac{1}{4}} e^{-\frac{(x-x_0)^2}{4 \sigma_h^2}}
\end{equation}
with $\sigma_h=\sqrt{\frac{\hbar}{2 m \omega}} $ and $x_0 \gg \sigma_h $. It shows~\cite{Schrodinger1926} that this initial waves packet corresponds to a small number of proper functions $ \varphi_n$ of the harmonic oscillator around the value $n\sim \frac{1}{2}(\frac{x_0}{\sigma_h} )^2$. We then obtain the coherent state:
\begin{equation}\label{eq:coht}
\Psi^h(x,t) = (2\pi \sigma_h)^{-\frac{1}{4}} e^{-\frac{(x-x(t))^2}{4 \sigma_h^2}+i\frac{m v(t) x}{\hbar}}
\end{equation}
where $x(t)=x_0 \cos\omega t$ and $v(t)=- x_0 \omega \sin \omega t$
correspond to the position and velocity of the centre of mass of a classical particle. The coherent state is therefore based on two scales, one of the classical type with $x_0$ and one of the quantum type with $\sigma_h$, which corresponds to the size of the wave packet and oscillates with its center of mass without changing shape. It is recalled that for the coherent state of a two-dimensional harmonic oscillator, the quasi-classical trajectory is an ellipse.

The extension of this interpretation to the electron in the hydrogen atom is based on Schrödinger's 1926 conjecture~:

\begin{Conjecture}
\textbf{ - Schrödinger's conjecture}: 
\textit{``It is certain that we can build wave packets gravitating on Kepler ellipses at a large number of quanta and forming the wave image of the electron of a hydrogen atom``.}~\cite{Schrodinger1926}
\end{Conjecture}

We will see that this conjecture can be based on the analogy with the coherent states of the harmonic oscillator for the wave function of an electron in a Rydberg state. 
We still give ourselves two more scales, the Bohr radius $a_0$ and a very large radius $a \gg a_0$. We then look for a wave packet corresponding to a small number of proper functions $\Psi_{n,l,m}(r,\theta,\varphi) e^{-i\frac{E_n t}{\hbar}} $ of the hydrogen atom around the value $n \sim \sqrt{\frac{a}{a_0}}$.

It was not until 1995 that Schrödinger's 1927 prediction was fulfilled for an electron in a Rydberg state.
We find in Bialynicki-Birula~\cite{Bialynicki1994},
Kalinski, Eberly, Buchleitner and Delande~\cite{Buchleitner1995} the
first constructions of a non-dispersive wave packet in
dimension 3 for the hydrogen atom in the presence of a microwave
field of circular polarized.\textit{''By passing through the rotating reference frame
the system becomes independent of time and a \textbf{stable} 
fixed point is located at a finite distance from the nucleus along the axis of the microwave electrical field. In the
laboratory reference frame, it corresponds to a
Bohr circular orbit passed throught at a constant angular velocity
equal to that of the microwave; this orbit is the centre of
the resonance island. In its vicinity, it is possible to build
quantum wave packets that rotate around the nucleus without being
distorted. These wave packets are not Gaussian, but
totally non-dispersive.''}~\cite{Delande2005}.
These non-dispersive wave packets correspond to periodic trajectories and are eigenvectors of the
Floquet operator.

\begin{figure}[H]
\begin{center}
\includegraphics[width=0.4\linewidth]{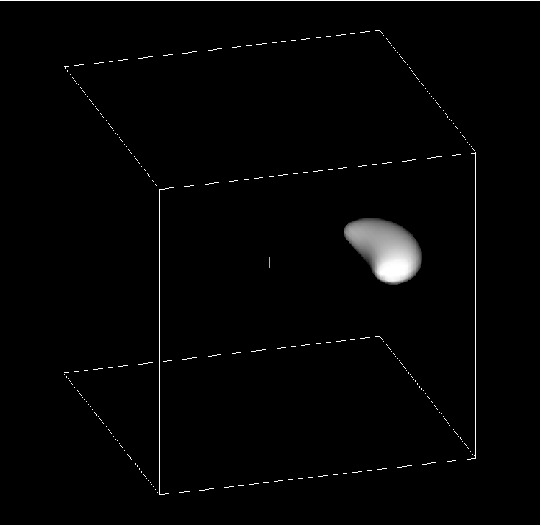}
\caption{\label{fig:delande3} Coherent wave packet of hydrogen atom, from the image of~\cite{Buchleitner1995}.}
\end{center}
\end{figure}
We note that these wave packets do not correspond to the usual solutions of quantum mechanics textbooks, which are stationary solutions of the electron and not wave packets located on a periodic trajectory.

Figure~\ref{fig:delande3} shows such a packet of waves, in
banana shape, calculated in a frequency field of about $30~Ghz$
 with a main quantum number $n=60$. The package is at
about 4,000 Bohr radius of the nucleus and revolves around it
in the horizontal plane without deforming.

Non-dispersive waves were successfully formed in experimental conditions for th first time in 2004
by Maeda and Gallagher~\cite{Maeda2004}, with
the observation of a lifetime greater than 15,000 periods of the
field, compared to the 10 periods observed in the absence of a field
for the dispersion of the wave packet.

\begin{Remarque}\textbf{- Wave packets with or without a periodic field -}
The numerically simulated wave packets in 1995 and those made experimentally in 2004 are created and maintained thanks to the presence of a periodic external field. One could therefore refute the conclusion of the existence of periodic wave packets without the presence of this external field. However, as Maeda and Gallagher point out, the field has no influence on the existence of periodic wave packets without deformation, but only on the number of periods without dispersion. 
\end{Remarque}

As Schrödinger had noted, these dynamic wave packets are difficult to determine because there is no analytical representation of them. 
But we can analytically calculate the dynamics of the center of mass of these states using the Floquet and Ehrenfest theorems.

Thus, for a bound particle such as the electron in the hydrogen atom, Schrödinger's interpretation restricted to the internal wave function is considered valid:
it is as if the electron were an electronic cloud with a charge density of $\rho_{int}=e |\varphi(\textbf{x},t)|^2$.
This interpretation of the internal wave function is extended to free particles.

This is the conclusion Schrödinger drew at the Solvay congress in 1927:

\textit{``I found the following way of looking aat things useful; it may be a little naïve but it is easy to grasp.
The classical system of material points does not really exist, but there is something that continuously fills all the space [...] the real system is a composite image of the classical system in all its possible states, obtained by using $\phi \phi^*$ as a "weight function".
The systems to which the theory is applied are classically composed of a large number of charged material points. As we have just seen, the charge of each of these points is distributed continuously through space and each charge point e provides the contribution of the $e \int \phi  \phi^* dx dy dz$ to the charge of the quarterly volume element $dx dy dz$. 
As $\phi  \phi^*$ generally depends on time, these charges vary''}.

\subsection{Schrödinger's generalized interpretation of the internal wave function}

Let us consider a quantum system composed of $N$ particles. At the initial moment of its preparation, its total wave function $\Psi_0$ emerges, as we saw in remark~\ref{rmq:emergence} of section~\ref{sect:decomposition}, from the entanglement of the initial functions $\Psi_0^j$ of the particles $j$ from which it is composed:
\begin{equation}\label{eq:fo}
\Psi_0 (\textbf{x}_1,\textbf{x}_2,\textbf{x}_3,...,\textbf{x}_N)= F(\Psi^1_0(\textbf{x}_1), \Psi^2_0(\textbf{x}_2), \Psi^3_0(\textbf{x}_3),...,\Psi^N_0(\textbf{x}_N)).
\end{equation}
For the EPR-B experiment, this equation corresponds to equation (\ref{eq:singuletEPR}) using Pauli's antisymmetrization to obtain the singulet wave function as we recalled in section~\ref{sect:EPR-B}

The evolution of the quantum system is then given by Schrödinger's equation which calculates the total wave function $ \Psi(\textbf{x}_1,\textbf{x}_2,\textbf{x}_3,...., \textbf{x}_N, t)$ from its initial condition $\Psi_0 (\textbf{x}_1,\textbf{x}_2,\textbf{x}_3,...,\textbf{x}_N) $.

To generalize Schrödinger's interpretation to cases of a quantum system of several particles, we rely on the following conjecture~: 

\begin{Conjecture}
\textbf{ - Schrödinger's generalized conjecture}: 
\textit{If the $N$ initial wave functions $ \Psi^j_0(\textbf{x}_j)$ satisfies equation (\ref{eq:fo}), then we can uniquely deduce  $N$ individual time-dependent wave functions $\Psi^j(\textbf{x}_j,t) $ from the
total wave function $ \Psi(\textbf{x}_1,\textbf{x}_2,\textbf{x}_3,....,\textbf{x}_N,t)$, which satisfies:}
\begin{equation}\label{eq:ft}
\Psi(\textbf{x}_1,\textbf{x}_2,\textbf{x}_3,...,\textbf{x}_N,t)= F_t(\Psi^1(\textbf{x}_1,t), \Psi^2(\textbf{x}_2,t), \Psi^3(\textbf{x}_3,t),...,\Psi^N(x_N,t)).
\end{equation} 
\end{Conjecture}

This possibility of retroengineering is a strong hypothesis. However, we have shown for the spinners in the EPR-B experiment that this type of re-creation is possible.

Within the framework of this conjecture, Schrödinger's interpretation of the internal wave function is then deterministic. In addition, the position of the center of mass of the quantum system particle $j$ is equal to:
\begin{equation}\label{eq:vitessecg}
X^j(t)=\int \textbf{x } |\Psi^j(\textbf{x},t)|^2 d\textbf{x}
\end{equation}
and verifies the Ehrenfest theorem.

\begin{figure}[H]
\begin{center}
\includegraphics[width=0.6\linewidth]{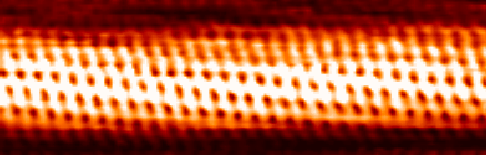}
\caption{\label{fig:silicon} 
Carbon nanotube observed with a tunneling microscope (STM) [source: Taner Yildirim (The National Institute of Standards and Technology - NIST) - Public Domain]}
\end{center}
\end{figure}

Figure~\ref{fig:silicon} is a topography of a carbon nanotube observed with a tunneling microscope; This image can be considered as a 2D representation of the standard of the nanotube's internal wave function.

We now provide some additional arguments in favour of this generalized Schrödinger interpretation for the internal wave function:

\begin{itemize}
 \item This is the simplest realistic interpretation for the internal wave function. The usual criticism (the measurement problem) is no longer valid because it is only intended for the interpretation of the external wave function.
 \item This interpretation is the basis of the \textbf{model of the elastically bound electron}.
 \item The first theorem of Hohenberg and Kohn~\cite{Hohenberg1964}, which is at the basis of \textbf{the theory of functional density} and which stipulates that a given electron density corresponds to a single wave function, is basically compatible with Schrödinger's interpretation of the internal wave function. The kinetic energy of electrons is then approximated as an explicit functional of density, while the contributions of core-electron attraction and electron-electron repulsion are treated in a classical way.
 \item The theory of the double solution with Schrödinger's interpretation is therefore also fundamentally compatible with the methodology of \textbf{molecular dynamic}.
 \item Let us finish with the recent experiment (2019) by Minev, Devoret et al.~\cite{Minev2019} on \textit{``the jump from the fundamental state to an excited state of a three-level superconducting artificial atom''}. She seems to agree with Schrödinger in his 1952 discussion with the Copenhagen-Göttingen school on quantum jumps, because \textit{``the experimental results show that the evolution of each jump performed is continuous, coherent and deterministic''}.
As Devoret explains: \textit{``Our experimental results show that quantum jumps are unpredictable and discrete (as Bohr thought) over long periods of time, they can be continuous (as suggested by Schrödinger) and predictable for a short period of time''}. Because, just before a jump occurs, there is always a latency period (a few microseconds) during which it is possible to acquire a signal that alerts you to the next jump.
These continuous and deterministic transitions are consistent with Schrödinger's generalized interpretation. The same should apply to operators of continuous and deterministic creation and annihilation.
\end{itemize}
\begin{Remarque}\textbf{- Neglected external wave function -}
The state transitions of the internal wave function are measured indirectly via the energy of the emitted particles (photons, electrons). 
This is the case for spectral measurements of atomic vapour lines (e.g. Balmer lines or Franck and Hertz experiments) or Dehmelt quantum jump experiments (fluorescence on a single three-level ion) and Minev and Devoret.
In these experiments, the external wave function (of the centre of mass) of the quantum system is neglected.
This is no longer the case in atom cooling experiments, where the recoil of the center of mass after each absorption/emission must be taken into account.
\end{Remarque}

The above arguments are not sufficient to exclude all other interpretations. 
For example, the interpretation of dBB for the internal wave function is the most plausible realistic and deterministic alternative. 
It also remains in continuity with the interpretation of the external wave function.

\begin{Remarque}\textbf{- The interpretation of dBB of the internal wave function -}
Schrödinger's interpretation, which we have defended, is based on the existence of non-dispersive wave packets whose evolutions are deterministic. The interpretation of dBB also applies very well to non-dispersive wave packets. 
Let's show it on the case of the coherent state of a harmonic oscillator. From the equation (\ref{eq:coht}) we derive that the velocity of dBB is $v^h (x,t)=\frac{\triangledown S^h(x,t)}{m}=v(t)$. For an initial position particle $x_0 + \eta$, where $ \eta$ is randomly drawn in a Gaussian of standard deviation $\sigma_h$, the dBB trajectory is $x_{\eta}(t)= x(t) +\eta$. 
When we tend $\hbar $ towards 0, $x_{\eta}(t)$ also tends towards the classical trajectory $x(t) $ and the interpretation of dBB is coherent to represent a single particle.
The same is true for Schrödinger's generalized interpretation as Norsen et al. recently showed \cite{Norsen2014} : for particles without non-relativistic spin, it is also possible in dBB's pilot wave theory to replace the wave function $\Psi(\textbf{x}_1,\textbf{x}_2,\textbf{x}_3,...., \textbf{x}_N,t)$ in the configuration space by N wave functions $\Psi^j(\textbf{x}_j,t) $ in the 3D physical space. These wave functions are the N conditional wave functions introduced by Dürr, Goldstein and Zanghi \cite{Durr2004}:
\begin{equation*}
\Psi^j(x_j,t)= \Psi(x_1, x_2,...,x_N,t)\vert_{x_i=X_i(t)~~for~~~i\neq j}
\end{equation*}
where $X_i(t)$ is the position of the particle $i$ in Bohmian mechanics.
\end{Remarque}

Finally, it should be noted that it is also possible to restrict the Copenhagen interpretation to the internal wave function.
 
\begin{Remarque}\textbf{- Copenhagen interpretation restricted to the internal wave function -}
Note that if we restrict the Copenhagen interpretation to the internal wave function alone, it becomes coherent again.
Indeed, the major criticism of the Copenhagen interpretation concerns neither realism nor determinism, but the incoherence of having a postulate at the time of measurement (postulate of reduction of the wave packet) that is not explicitly related to the evolution of the wave function (Schrödinger equation). 
Quantum theory does not specify whether to use the wave packet reduction assumption or the Schrödinger equation.
This incoherence leads to the so-called problem of measurement~\cite{Maudlin1995} and constitutes the major theoretical problem of quantum mechanics. 
However, if the Copenhagen interpretation concerns only the internal wave function, then the assumption of a reduction of the wave packet is no longer necessary. 
Indeed, the deterministic measurement is supported by the external wave and the measurement problem does not arise for the internal wave function.
However, the recent experience of Minev, Devoret and al.~\cite{Minev2019} seems to invalidate this interpretation even for the internal wave function.
\end{Remarque}

\section{A crucial experiment: asymmetric interference}
\label{sect:experienceCruciale}

To define an experiment to validate the simultaneous existence of external and internal wave functions, we will use the results of two experiments, an already old experiment (1983) by Serge Haroche's team~\cite{Fabre1983} with Rydberg atoms  
 and a more recent experiment (1999) by Zeilinger's team~\cite{Arndt1999} with  $C_{60}$ molecules.

In the Haroche experiment, we measure \textit{``the transmission of a jet of Rydberg atoms through a metal grid made of a network of micrometric slits. The transmission decreases linearly with the square of the main quantum number $n$, with a break for a maximum value of $n$''.} So as soon as $n$ is large enough ($n\geq60$ in this experiment for slits with a width of $1~\mu m$), the particle density measured after the slits is zero. The usual interpretation, which makes no difference between the wave and the particle, assumes that, if the particles do not pass, the (total) wave function of the particle does not pass either and is therefore null behind the slots; this is the standard assumption.

In the Zeilinger interference experiment with $C_{60}$ molecules, the experimental results show that the density of the $C_{60}$ molecules after the slits is calculated by taking only the wave function of the mass center (external wave) and ignoring the internal wave of the molecule.

In the interpretation of the external and internal functions in which the center of mass wave and the molecule wave are differentiated, the conditions for passing through a slit of the external wave and the internal wave may be different: if a slit is smaller than the size of the internal wave of the particle, then the particle will not pass through the slit and nor will its externe wave. But if there are two slots, one smaller than the size of the internal wave and one larger, the internal wave can pass through the large slit but not through the small one. If it goes through the big one, then what about its external wave~? It will obviously pass through the big slit, but will it also pass through the small one~? 
This is the alternative hypothesis.

We proposed two experiments to test this hypothesis, one using fullerene molecules~\cite{Gondran2006a}, the other using Rydberg atoms~\cite{Gondran2008b} and which is currently feasible, unlike the one with fullerenes.

The idea of these experiments is to generete interferences, not from two identical slits, but from a slit A (large compared to the size of a Rydberg atom or a fullerene) and a grid B of several hundred small slits that do not let the atom or molecule through.  
For example, for Rydberg atoms with $n=60$, a slit A of 100 $\mu$m and a grid B of 1000 slits of 0.1                         
$\mu$m. Simulation shows~\cite{Gondran2008b} that the interference results can be used to discriminate between the standard and alternative hypothesis.

\section{Conclusion}
\label{sect:conclusion}

We proposed, in the pre-relativist context, a theory of the double solution that responds to Louis de Broglie's specifications~\cite{deBroglie1971}:  
\begin{quotation}
\textit{``I was introducing,
under the name of "double-solution theory" the idea
that we had to distinguish between two solutions that are distinct but
intimately related to the wave equation, one that I called
the $u$ wave, being a real and non-standard physical wave
with a local accident defining the particle and
represented by one singularity, the other, Schrödinger's $\Psi$ wave , normalizable and devoid of singularity, which would only be
a representation of probabilities.''}~\cite{deBroglie1971}
\end{quotation}
In this theory, the total wave function of a quantum system is decomposed into an external wave function corresponding to the external variables of the quantum system (position and velocity of the center of mass, orientation of the global spin) and an internal wave function corresponding to the internal variables of the quantum system. The internal and external wave functions correspond respectively to the u and $\psi$ waves of the Broglie specification, the total wave being approximately the product of the u and $\psi$ waves. The internal wave function u is a physical wave representing the microscopic quantum system (size $\sim $ h) while the external wave function $\psi$ is a wave associated with the center of mass of the quantum system.

By studying the convergence of the external wave function $\psi$ when h tends towards 0, we have shown that it is very likely to "pilot" the center of mass of the quantum system as in the interpretation of the Broglie-Bohm pilot wave. It is also compatible with the Born statistical interpretation.

The interpretation of the internal wave function $u$ can be related to Schrödinger's interpretation where the densities of the particles wave functions correspond to real densities. 
The particles center of mass follow trajectories from the Ehrenfest theorem and this interpretation of the internal wave function is deterministic. We conclude that our approach to the double solution is deterministic. However, dBB interpretation is also possible. It should be noted that in both cases, our approach to the double solution is realistic and deterministic.

This double solution explained by the external and internal wave functions is a new interpretative framework to understand the debates on the interpretation of quantum mechanics, in particular those of the Solvay congress of 1927. Each of the founding fathers had a share of the truth about one of the two wave functions they generalize to the global wave function: de Broglie for the pilot wave of the external wave function, Schrödinger for the soliton of the internal wave function, Born for the statistical interpretation of the external wave function and the Copenhagen school for the complete description of the internal wave function.
This decomposition gives a simple explanation of wave-corpuscle duality; the external wave drives the system whereas the internal wave represents the corpuscular part of the system.

This decomposition also makes it possible to see under a new paradigm the relationships between quantum mechanics and general relativity, gravitation only appearing in the external wave function.

However, there remains much work to do on this theory of the double solution. In this respect, Louis de Broglie's comments in 1966 on the direction in which research would continue and the way in which he would be viewed by history are both poignant and prescient:
\begin{quotation}
\textit{``The future, a future I probably won't see, may decide
the question: it will say whether my current point of view is 
of an already quite old man who remains attached to the ideas of his youth or, if it reflects the foresight of a researcher who has been refleting his whole life long on the most fundamental problem of 
contemporary physics''}~\cite{deBroglie1966}
\end{quotation}

\bibliographystyle{abbrv}
\bibliography{/home/gondran/Documents/mq/biblio/biblio_mq}

\end{document}